\documentclass[12pt]{iopart}
\usepackage{epsfig,graphicx}
\usepackage[colorlinks=true,linkcolor=blue,citecolor=blue]{hyperref}
\usepackage{iopams}

\usepackage{color} 
\newcommand{\adag}{a^{\dagger}}
\newcommand{\adaga}{a^{\dagger}a}
\newcommand\ket[1]{\left|\textstyle{#1}\right\rangle}

\newcommand\braket[1]{\left\langle\textstyle{#1}\right\rangle}
\newcommand{\upx}{\ket{\uparrow}_x}
\newcommand{\downx}{\ket{\downarrow}_x}

\newcommand{\up}{\ket{\uparrow}}

\newcommand{\Hz}{\textrm{Hz}}
\newcommand{\kHz}{\textrm{kHz}}
\newcommand{\MHz}{\textrm{MHz}}

\begin{document}
\date{\today}
\title[]{A robust scheme for the implementation of the quantum Rabi model in trapped ions}
\author{Ricardo Puebla, Jorge Casanova and Martin B. Plenio}
\address{Institut f\"{u}r Theoretische Physik and IQST, Albert-Einstein Allee 11, Universit\"{a}t Ulm, 89069 Ulm, Germany}
\ead{ricardo.puebla@uni-ulm.de}
\begin{abstract}
We show that the technique known as concatenated continuous dynamical decoupling (CCD) can be applied to a trapped-ion setup for a robust implementation of the  quantum Rabi model in a variety of  parameter regimes. These include the case where the Dirac equation emerges, and the limit  in which a quantum phase transition takes place. We discuss the applicability of the CCD scheme in terms of the fidelity between different initial states evolving under an ideal  quantum Rabi model and their corresponding trapped-ion realization, and demonstrate the effectiveness of noise suppression of our method. 
\end{abstract}


\maketitle

\section{Introduction}
Quantum coherence is an essential prerequisite to observe and exploit the intriguing phenomena in the quantum realm~\cite{Nielsen00}. Indeed, technologies relying on those quantum properties are expected to surpass their classical counterparts in efficiency and performance. This new generation of quantum technologies encompasses a large diversity of possible applications which inlcude quantum simulation~\cite{Feynman82}, quantum metrology~\cite{Giovannetti04}, quantum communication~\cite{Gisin07} and quantum sensing~\cite{Wu16}, all of them requiring the preservation of quantum coherence for their correct functioning. In this respect, the loss of quantum coherence, or simply decoherence, is a crucial limitation  as it occurs due to the unavoidable interaction of the quantum system with an uncontrolled environment as well as to the presence of experimental imperfections. Hence, the long-time maintenance of the quantum coherence of an evolving system is highly desired although its realization constitutes a formidable task.

During the past decades considerable efforts have been invested in the development of theoretical schemes to circumvent, as much as possible, the effect of the noise in the system with the  goal of prolonging coherence times. Among them we find techniques such as decoherence-free subspaces~\cite{Lidar13}, quantum error correction~\cite{Lidarbook13}, or dynamical decoupling~\cite{Souza12}. These are methods designed to handle specific noise scenarios, and present different benefits concerning noise supression. In particular, dynamical decoupling constitutes a promising tool to handle non-Markovian noise,  and it is the central object of study in this article. In its continuous wave configuration, the effect of dynamical decoupling corresponds to the creation of a dressed basis with an energy gap such that, under certain circumstances that will be later developed, the effect of noise is suppressed. In addition, this technique allows for a {\em concatenated} configuration known as \textit{concatenated continuous decoupling} (CCD)~\cite{Cai:12} that consists in applying concurrently different driving fields to eliminate further sources of noise, including those from imperfect driving fields themselves.  Standard dynamical decoupling has been theoretically proposed in its continuous~\cite{Bermudez:12, Lemmer:13njp,Cohen15, Mikelsons15} and pulsed~\cite{Souza12, Carr54, Meiboom58, Casanova15} configurations. Furthermore, these techniques have already been used in both radio frequency and Penning traps in ~\cite{Timonei11, Tan13} (continuous case) and in~\cite{UyS09, Biercuk09, Biercuk09bis, Biercuk09bisbis} (pulsed case) as a method to suppress  noises on the registers and  to  drive robust single- and two-qubit gates. Furthermore, dynamical decoupling has been used to explore different models involving spin-spin interactions~\cite{Cohen15bis}. On the other hand, the CCD scheme has experimentally demonstrated its feasibility to preserve the coherence of an isolated nitrogen-vacancy center in diamond~\cite{Cai:12}. However, the convenience and possible benefits of the  CCD method  in an ion trap platform for quantum simulation purposes has not been proven yet.

In the present article we show how to apply the CCD scheme in a trapped-ion setting for a robust implementation of the paradigmatic quantum Rabi model that describes the interaction between a two-level system and one bosonic field mode. Despite of its apparent simplicity, this model exhibits a rich variety of physics, ranging from the relativistic Dirac equation~\cite{Lamata07, Gerritsma09, Casanova10r, Gerritsma11} to critical phenomena as it can undergo a second-order quantum phase transition~\cite{Hwang:15,Puebla:16}. We demonstrate that, within the CCD scheme, high fidelities can be achieved and maintained during long evolution times in an ion trap setup in the presence of different noise sources and realistic conditions. While an experimental verification of such scheme in an ion trap is still required, the present theoretical results are promising and open the door to the study of robust and noise-resilient trapped-ion quantum simulations.

We exemplify and support by means of detailed numerics the applicability of the CCD scheme realizing the quantum Rabi model in three different parameter regimes. First, the  case where the energy splitting of the two-level system matches the motional frequency  and the rotating-wave approximation can be applied. In this situation  the Jaynes-Cummings model~\cite{Jaynes63} emerges and we can observe Rabi oscillations. Second, the realization of the Dirac equation~\cite{Lamata07, Gerritsma09, Casanova10r, Gerritsma11} whose main hallmark is the Zitterbewegung, and finally, the extreme parameter regime~\cite{Casanova10} required to witness critical dynamics as a consequence of the emergence of a second-order quantum phase transition in the limit of strong coupling~\cite{Hwang:15,Puebla:16}. Additionally, we discuss possible drawbacks in the CCD scheme and identify particular situations where the method does not lead to an improved performance.

The present article is organized as follows. In Sec.~\ref{sec:OU} we introduce the Orstein-Uhlenbeck stochastic process~\cite{Orstein:30,Wang:45}, which we will use to model fluctuations in the trapped-ion setting as well as of the externally applied control fields. In Sec.~\ref{sec:ccd} the CCD scheme is presented and  explained. Furthermore, we show how CCD adapts to trapped-ion Hamiltonians giving rise to a noise protected quantum Rabi model in  Sec.~\ref{sec:TI}, while specific examples and their numerical simulations are shown in Sec.~\ref{sec:num}. Finally, we summarize the main conclusions in Sec.~\ref{sec:conc}.

\section{Stochastic fluctuations: Orstein-Uhlenbeck process}
\label{sec:OU}
A quantum system looses its quantum coherence due to an uncontrolled interaction with the environment. Such interaction introduces a stochastic noise or fluctuation in the system that we will model as an Orstein-Uhlenbeck (OU) stochastic process~\cite{Orstein:30,Wang:45,Gillespie:96}. This effective description successfully reproduces the exponential decay of the quantum coherence due to dephasing noise as measured by Ramsey interferometry~\cite{Wineland:98}, as well as the behavior of a quantum system under  fluctuations on the intensity of the applied radiation~\cite{Cai:12}. Moreover, as we will see later on, it also allows to vary the width of the spectral density, which quantifies the amount of power per unit of frequency. In this manner the OU process can describe different noise scenarios, and thus, it has been extensively used in the literature~\cite{Bermudez:12,Lemmer:13njp,Bermudez:13,Lemmer:13}.

 An OU process is characterized by two parameters, namely, $\tau$ and $c$, relaxation or correlation time and diffusion constant, respectively. While the former fixes the time in which the noise is correlated, the latter is proportional to the noise amplitude. A stochastic variable $X(t)$ that obeys an OU process has an exact update formula~\cite{Gillespie:96},
\begin{equation}
\label{eq:OU}
X(t+\Delta t)=X(t)e^{-\Delta t/\tau}+\left[\frac{c\tau}{2}\left(1-e^{-2\Delta t/\tau}\right)\right]^{1/2}N(t),
\end{equation}
for an arbitrary value of $\Delta t$. The term $N(t)$ stands for a temporally uncorrelated normally distributed random variable, i.e., $\overline{N(t)}=0$ and $\overline{N(t)N(t')}=\delta(t-t')$, where the overline denotes the stochastic average. The OU process is Gaussian, and hence, fully determined by its first and second moments, 
\begin{eqnarray}
\overline{X(t)}&=0 \\
\sigma^2[X(t)]&=\frac{c\tau}{2}\left(1-e^{-2t/\tau}\right),
\end{eqnarray}
where $\sigma^2[X]$ denotes the variance of $X$, and thus, $\sigma[X]$ its standard deviation.
 The power spectrum or spectral density, $S_X(f)$,  characterizes the nature of the noise, since it measures the amount of power per unit of frequency of $X(t)$ at a frequency $f$. The stochastic variable $X(t)$ can be written in Fourier series as $X(t)=\sum_n P_n e^{2\pi i f_n t}$ for $t\in[0,T]$ where $P_n$ are the corresponding Fourier coefficients at frequency $f_n$. Then, the spectral density can be defined in the $T\rightarrow \infty$ limit, as shown in~\cite{Wang:45}, as $S_X(f_n)=\lim_{T\rightarrow \infty} \frac{1}{T}\left|P_n \right|^2$. The spectral density will be of importance in the next section, Sec.~\ref{sec:ccd},  for the understanding of the noise decoupling  efficiency of the CCD method.  Indeed, for the particular case of an OU process,  $S_X(f)$ can be analytically calculated  giving rise to~\cite{Wang:45}
\begin{equation}
\label{eq:Sxf}
S_X(f)=\frac{c\tau^2}{1+4\pi^2\tau^2f^2}.
\end{equation}
Therefore, the relaxation time $\tau$ sets a boundary in the frequency domain between {\em white} noise, i.e. $S_X(f)\propto f^0$, and {\em Brownian} or {\em red} noise, i.e. $S_X(f)\propto f^{-2}$. This \textit{crossover} frequency $f_{cr}$ can be estimated as $S_X(f_{cr})/S_X(0)=1/2$, that is, $f_{cr}=1/(2\pi\tau)$.
\begin{figure}
\centering
\includegraphics[width=0.5\linewidth,angle=-90]{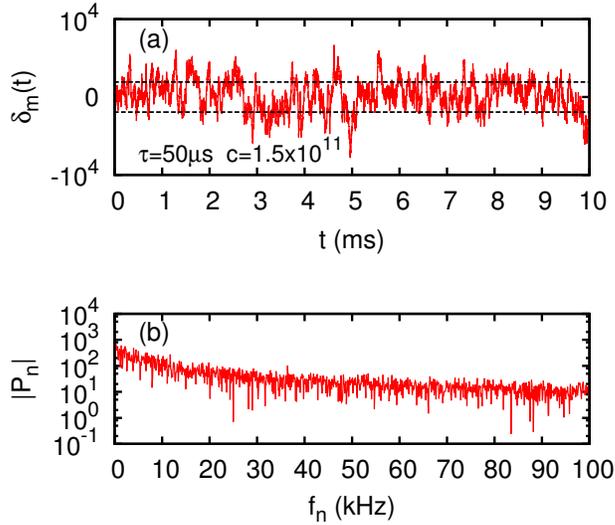}
\caption{{\small (a) Trajectory of a stochastic variable $\delta_m(t)$ obeying a OU process with $\tau=50 \ \mu$s and $c=1.5\times 10^{11} \ s^{-3}$. The dashed lines demarcate the area within the standard deviation $\pm\sigma[\delta_m(t)]$. (b) Coefficients $|P_n|$ as a function of the frequency $f_n$ of the Fourier transform of the stochastic variable $\delta_m(t)$.}}
\label{fig:FFT}
\end{figure}
In Fig.~\ref{fig:FFT} we show a typical trajectory of an OU process  for a fluctuating variable $\delta_m(t)$ and its Fourier transform. Note that $S_X(f_n)\propto \left|P_n\right|^2$.

\begin{figure}
\centering
\includegraphics[width=0.4\linewidth,angle=-90]{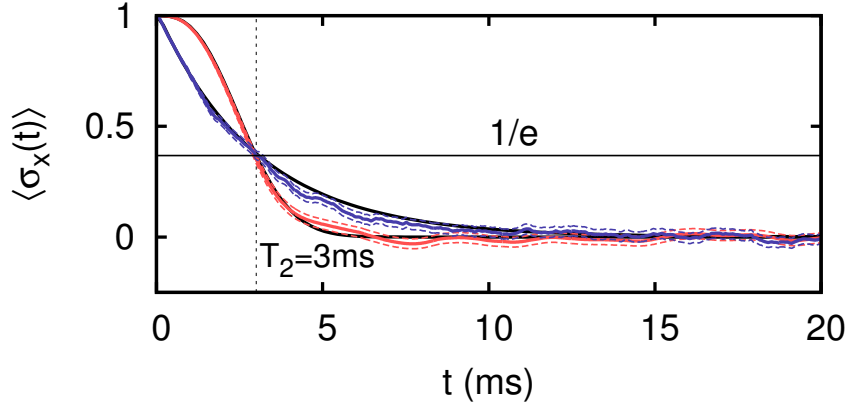}
\caption{{\small Decoherence due to a fluctuating $\sigma_z$ term (dephasing noise), which follows a OU process.  Two different noises have been considered: fast or short memory noise (solid-blue) $\tau=50 \ \mu$s and slow noise (solid-red) $\tau=5$ ms. The coherence time $T_2=3$ ms is fulfilled in both cases when the diffusion constant $c$ is calculated from Eq.~(\ref{eq:c}). An initial state $\upx$ is prepared, and then the expectation value $\braket{\sigma_x(t)}$ is calculated as an average over $1000$ stochastic trajectories, and presented  together with their corresponding sample variance (dashed lines). The solid black lines show the exact $\braket{\sigma_x(t)}$ according to Eq.~(\ref{eq:sx}). As expected, for slow noise  a Gaussian behavior is observed $e^{-(t/T_2)^2}$, while the case with fast noise decays exponentially $e^{-t/T_2}$.}}
\label{fig:coherences}
\end{figure}

Here we are interested in magnetic-field fluctuations or simply dephasing noise, which can be written as $H=\delta_m(t)/2 \ \sigma_z$ where $\delta_m(t)$ follows  Eq.~(\ref{eq:OU}). The coherence time of the system depends then on the properties of $\delta_m(t)$. For example, consider an initial state $\upx$ at $t=0$, i.e. $\sigma_x\upx=+\upx$, evolving under $H=\delta_m(t)/2 \ \sigma_z$, then it is easy to prove that
\begin{equation}
\label{eq:sx}
\braket{\sigma_x(t)}=e^{-\frac{1}{2}\overline{\varphi^2(t)}},
\end{equation}
where $\varphi(t)=\int_0^t ds \ \delta_m(s)$ is the time integral of the stochastic variable $\delta_m(t)$ and $\overline{\varphi^2(t)}$ its autocorrelation function that can be written as~\cite{Gillespie:96}
\begin{equation}
\label{eq:phi2}
\overline{\varphi^2(t)}=c\tau^2\left[t-\tau\left(\frac{3}{2}-2e^{-t/\tau}+\frac{1}{2}e^{-2t/\tau} \right)\right].
\end{equation}
The coherence time $T_2$ is defined as the time instant at which $\braket{\sigma_x(T_2)}=e^{-1}$. Hence, from Eq.~(\ref{eq:phi2}) and~(\ref{eq:sx}) it follows that 
\begin{equation}
\label{eq:c}
c=\frac{4e^{2T_2/\tau}}{\tau^2\left(4e^{T_2/\tau}\tau-\tau+e^{2T_2/\tau}(2T_2-3\tau) \right)}
\end{equation}
that is, for a given $\tau$ and a coherence time $T_2$, the diffusion constant can be determined. Nevertheless, depending on whether the noise is fast, i.e. with short memory, meaning $\tau\ll T_2$, or slow, i.e. with long memory, which corresponds to $\tau\gtrsim T_2$, the coherence decays differently. Indeed, exponential decay is achieved when $\tau\ll T_2$ which is the typical scenario in ion traps~\cite{Wineland:98}.  In this case Eq.~(\ref{eq:c}) acquires a  simpler form: $T_2\approx2/c\tau^2$. In contrast, for slow noise a Gaussian decay is observed. In Fig.~\ref{fig:coherences} we plot  $\braket{\sigma_x(t)}$ as a function of the evolution time $t$ for an initial state $\upx$ evolved under fast and slow noise, considering $T_2=3$ ms, $\tau=50 \ \mu$s and $\tau=5$ ms, and $c$ obtained according to Eq.~(\ref{eq:c}).  We can observe how the numerical stochastic average $\braket{\sigma_x(t)}$ agrees with the exact expression in Eq.~(\ref{eq:sx}).

\section{Concatenated Continuous Decoupling (CCD)}
\label{sec:ccd}
In this section we explain the technique known as dynamical decoupling in a concatenated scheme (CCD)~\cite{Cai:12} that corresponds to the addition of several continuous decoupling fields. Note that the use of continuous fields, not pulsed,  will be maintained throughout the article. Consider a situation where the Hamiltonian is $H=\omega_0(t)/2\ \sigma_z$ where $\omega_0(t)=\omega_0+\delta_m(t)$ with $\delta_m(t)$ the  stochastic fluctuation of $\omega_0$, which strongly affects the quantum coherence  of the system. Then,  in order to eliminate its effects a  continuous  driving field with Rabi frequency $\Omega$ is introduced. This situation is described by the Hamiltonian
\begin{equation}
\label{eq:ccd1}
H=\frac{\omega_0}{2}\sigma_z+\frac{\delta_m(t)}{2}\sigma_z+\Omega\cos(\omega t)\sigma_x.
\end{equation}
In an interaction picture w.r.t. $\omega_0/2\sigma_z$ we have
\begin{equation}
H^I=\frac{\delta_m(t)}{2}\sigma_z +\frac{\Omega}{2}\left[\sigma^+\left(e^{i(\omega_0+\omega)t}+e^{i(\omega_0-\omega)t}\right)+\textrm{H.c.}\right],
\end{equation}
thus, selecting $\omega=\omega_0$ and invoking the rotating-wave approximation (RWA), the previous Hamiltonian (in the case $\Omega\ll\omega_0$) reads
\begin{equation}
H^I\approx \frac{\delta_m(t)}{2}\sigma_z+\frac{\Omega}{2}\sigma_x.
\end{equation}
The first term on the r.h.s of the above equation  produces no transition in the basis $\left\{\upx,\downx \right\}$ as long as the fluctuating term, $\delta_m(t)$, has vanishing Fourier coefficients, $|P_n|\ll 1$, in the vicinity of frequencies $f_n\approx \Omega$. In other words, to protect the system against the noise, the Rabi frequency $\Omega$ must lie in the region in which the noise spectrum is negligible. In this manner, transitions in the dressed basis $\left\{\upx,\downx \right\}$ as a consequence of the stochastic term $\delta_m(t)/2 \ \sigma_z$ have an energy penalty and can be neglected. We will denote this first step as the {\em first layer} of protection, since only one additional driving has been introduced. From a more rigorous point of view, that noise elimination is achieved after the application of a RWA on each of the noise components as a consequence of the presence of the term $\frac{\Omega}{2} \sigma_x$. In addition, and because the RWA presents a slightly different behavior depending on the initial state of the system, the proposed method inherits its dependence. Note however the existence of  certain states for which its evolution under noise and the Hamiltonian just gives rise to a global phase. For such dark states, introducing a first layer deteriorates the coherent evolution since, in the rotated basis, noisy terms are able to produce transitions. In~\ref{subsec:QRM} we will comment more about this scenario and show an example. Now one should also consider that the Rabi frequency $\Omega$ is not completely stable and represents another source of fluctuations, that is, $\Omega \equiv \Omega [1 + \delta_\Omega(t) ]$ with $\delta_\Omega(t)$ another stochastic fluctuation with a small amplitude. However, the  CCD scheme offers the possibility  to further protect the system against  $\delta_\Omega(t)$ with a {\em second layer} by introducing one additional driving to cancel $\delta_\Omega(t)$~\cite{Cai:12}. 

\begin{figure}
\centering
\includegraphics[width=1\linewidth,angle=0]{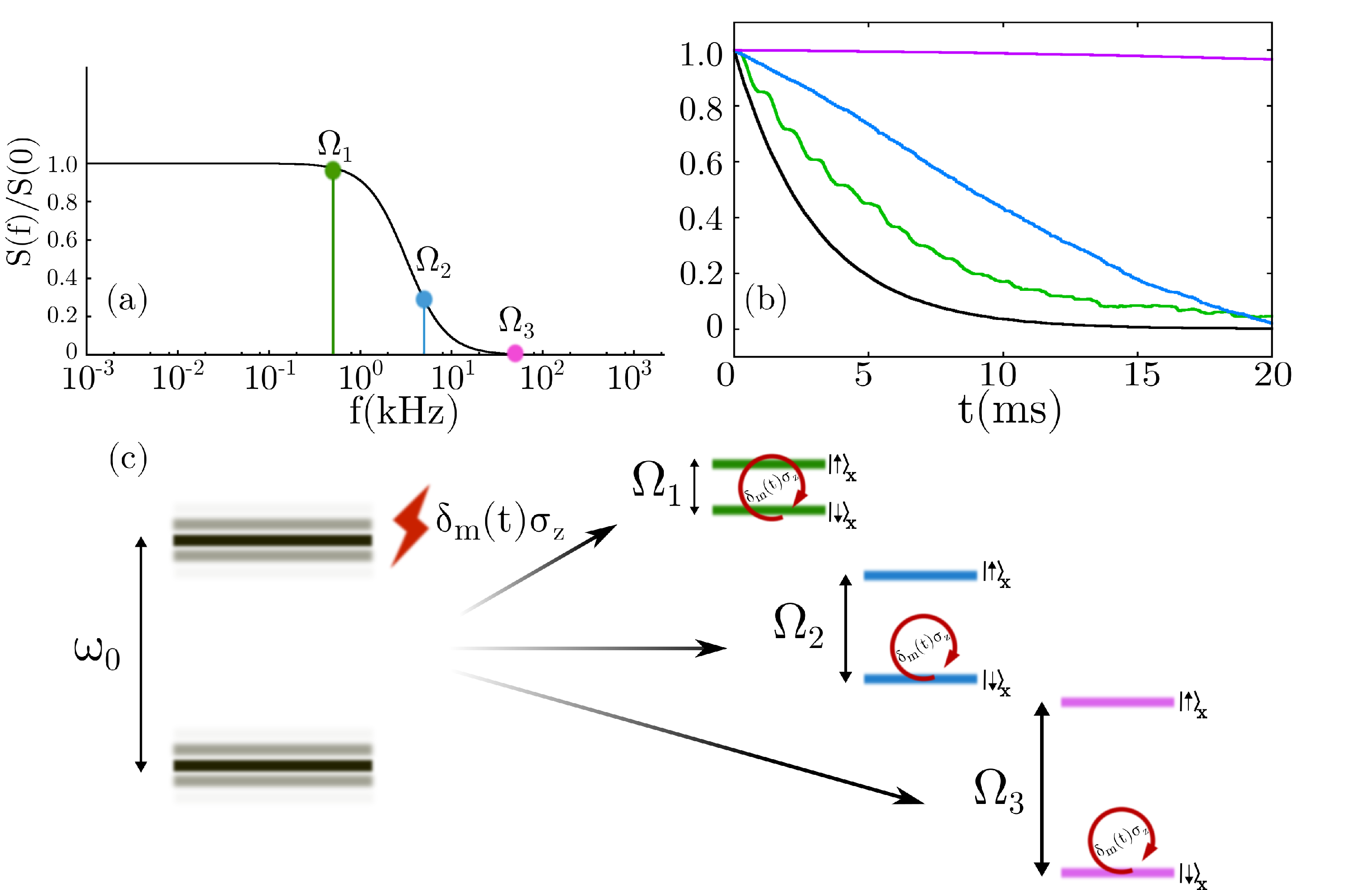}
\caption{{\small Schematic representation of the CCD scheme. In (a) the normalized power spectrum of the noise is plotted. Depending on the Rabi frequency $\Omega_i$ of the additional pulse, different evolution of the coherences is observed (b). As sketched in (c), the original basis suffers dephasing. Then, if the introduced $\Omega_i$ is small compared to the characteristic frequency of the noise, there is essentially no protection, while $\Omega_i\gtrsim f_{cr}$ coherence times are enhanced significantly as the noise term $\delta(t)\sigma_z$ is not enough to produce transitions in the new dressed basis. Noise parameters are $\tau=50 \ \mu$s and $T_2=3$ ms, while the Rabi frequencies $\Omega_1=2\pi\times0.5 \ \kHz$, $\Omega_2=2\pi\times 5 \ \kHz$ and $\Omega_3=2\pi\times50 \ \kHz$, and $\omega_0\gg\Omega_{3}$ such that a RWA can be safely applied.}}
\label{fig:schemeCCD}
\end{figure}

In Fig.~\ref{fig:schemeCCD} we sketch the main idea behind the effectiveness of dynamical decoupling to cancel interfering stochastic processes. In Fig.~\ref{fig:schemeCCD} (b) the evolution of the coherences as a function of the evolution time is plotted for three different drivings. The success depends on the properties of the noise (a): when the Rabi frequency of the driving does not exceed the crossover frequency of the noise ($\Omega_1<f_{cr}$) no protection is achieved. On the contrary, as the Rabi frequency gets larger, $\Omega_{2,3}\gtrsim f_{cr}$, the quantum coherence is preserved during longer times since transitions due to the original noise occur with a smaller probability in the new dressed basis. This shows the crucial interplay between noise properties and driving frequencies in a dynamical decoupling scheme. Then, one can apply the same criteria to cancel further fluctuations of additional drivings fields in the CCD scheme. Note that the same techniques can be applied to other noise models that present a similar behavior, i.e. models exhibiting a spectral density that vanishes for asymptotically large frequencies. 

\section{Trapped-ion Hamiltonian and CCD}
\label{sec:TI}

Consider a trapped-ion with internal electronic structure described by $\omega_I/2 \ \sigma_z$ and $\nu\adaga$ representing the motional mode energy with $\nu$ the trap frequency. The interaction created by a laser irradiation is captured in the term $\Omega_j/2\sigma_x\left[e^{i(k\hat{x}-\omega_jt-\phi_j)}+\textrm{H.c.}\right]$. Hence, under the influence of  applied radiation the trapped-ion Hamiltonian reads~\cite{Leibfried:03}
\begin{equation}
\label{eq:TIH}
H=\frac{\omega_I}{2}\sigma_z+\nu \adaga +\sum_j \frac{\Omega_j}{2}\sigma_x\left[e^{i(k_j\hat{x}-\omega_jt-\phi_j)} +\textrm{H.c.} \right].
\end{equation}
where $k_j$ is the wave vector of each laser field,  $\omega_j$ its frequency, $\phi_j$ an initial phase,  $\Omega_j$ the Rabi frequency of the $j$th laser, and  $\hat{x}$ the ion position  operator.

Before starting with further developments, let us introduce some typical values of the parameters in the previous equation according to the state-of-the-art in experiments with $^{40}\rm{Ca}^{+}$~\cite{Gerritsma09,Gerritsma11}. Here, the axial trap frequency is $\nu = 2 \pi \times 1.36$ MHz, $\omega_I$ is on the optical regime at $729$ nm, i.e. $\omega_I =2\pi \times 4\cdot 10^{14}$ Hz, and the Rabi frequency is typically on the order of several kHz~\cite{Gerritsma09,Gerritsma11}. Additionally, we should consider the coherence time of the internal  levels of the ions  as the main limiting factor that affects to the quality of the experiments with $^{40}\rm{Ca}^+$~\cite{Gerritsma09,Gerritsma11}. As we already commented this is caused by magnetic-field fluctuations which give rise to a coherence time $T_2 \approx 3$ ms, see~\cite{Gerritsma11}. We  will consider this value throughout the present article. Note however that, by using a cryogenic setup~\cite{Brandl:16}, a longer coherence time of $T_2\approx18$ ms has already been achieved. Additionally, laser-intensity fluctuations are present in any realistic ion trap experiment, while its frequency $\omega_j$ and phase $\phi_j$ can be very accurate. Although these magnetic and intensity fluctuations are the main limiting factor for the coherence time of the system, there are still another  sources of noise which will be not  considered here as they will produce significant effects only on time scales significantly longer than  $T_2=3$ ms. In this respect, phonon dephasing has been measured with an incidence of few $\Hz$~\cite{Kaler03}. This provides a limit of the time scale across which the dynamics can be observed, which is, approximately, two orders of magnitude larger than the one we could consider if the magnetic noise is not eliminated. Concerning the heating rate it can be estimated that, on average, one phonon is gathered in $\sim 100$ ms~\cite{Kaler03}, or  in $\sim 500$ ms for a cryogenic setup~\cite{Brandl:16}. Furthermore, the lifetime of the qubit for the D$_{5/2}$ state of $^{40}$Ca$^+$ is $\sim 1$s~\cite{Kaler03}. 

Regarding the trapped-ion Hamiltonian, in the interaction picture w.r.t. $H_0=\frac{\omega_0}{2}\sigma_z+\nu \adaga$, it reads
\begin{eqnarray}
\label{eq:Hi_orwa}
H^{I}&=e^{i(\frac{\omega_I}{2}\sigma_z+\nu \adaga)t}H_1e^{-i(\frac{\omega_I}{2}\sigma_z+\nu \adaga)t}\nonumber \\ &\approx \sum_j \frac{\Omega_j}{2}\left[\sigma^+e^{i\eta_j(ae^{-i\nu t}+ae^{i\nu t})}e^{i(\omega_I-\omega_j)t-i\phi_j}+\textrm{H.c.} \right],
\end{eqnarray}
where we have already performed the optical RWA, i.e., we neglect the terms that rotate at frequency $\omega_I+\omega_j$ (counter rotating terms). Since $\omega_j$ will be chosen such that $\omega_j\approx \omega_I$ and because $\Omega_j\ll \omega_I+\omega_j$, this approximation can be safely carried out.  We denote $\Delta_j=\omega_I-\omega_j$, thus, choosing $\Delta=0$, $\nu$ or $-\nu$ one arrives to a {\em carrier}, {\em red sideband} or {\em blue sideband} interaction, respectively, when the system is adjusted to lie within the Lamb-Dicke regime ($\eta_j\sqrt{ \braket{(a+\adag)^2} }\ll 1$). Here, the Lamb-Dicke parameter $\eta_j$ is $\eta_j=k_j x_0$ where $x_0=(2m\nu)^{-1/2}$, $m$ the mass of the ion and $\hbar=1$ throughout the whole article; thus, $\hat{x}=x_0\left(a+\adag\right)$. Finally, we would like to remark that all the numerical simulations of trapped-ion Hamiltonians presented in this article have been performed after the optical RWA and without further assumptions.

\subsection{CCD for a single trapped-ion setup}
\label{subsec:CCDRabi}
 We discuss now how to employ a CCD scheme in a single trapped-ion setup. In~\cite{Pedernales:15} it is demonstrated that, by using two traveling waves to excite the red- and blue sideband transitions, and by setting properly the parameters $\Omega_{1,2}$, $\phi_{1,2}$ and $\omega_{1,2}$, the Rabi model can be simulated in a variety of parameter regimes which includes the Dirac equation as a particular case. However,  the presence of different  noise sources could significantly deteriorate its realization. Therefore, a noise-resilient implementation is desired to enhance coherence control and fidelity. For that reason, in the following we apply a CCD scheme to a single trapped-ion setup. We use the {\em first layer} (\ref{subsubsec:1layer}) to tackle the dephasing noise as it is the main limiting factor for the coherence time of the system, while the {\em second layer} is introduced to handle laser-intensity fluctuations (\ref{subsubsec:2layer}).

\subsubsection{First layer}
\label{subsubsec:1layer}
In order to achieve the Rabi model within the CCD scheme, we apply an extra laser, denoted by the subscript $a$, with the objective to introduce a  term $\Omega_a \cos(\omega_I t)\sigma_x$ into the dynamics. This is accomplished by setting $\omega_a=\omega_I$ (resonant with the frequency splitting of the ion), $\phi_a=0$ and a Rabi frequency $\Omega_a\ll \omega_I$. Then, the trapped-ion Hamiltonian in a rotating frame w.r.t. $H_0=\omega_I/2 \sigma_z +\nu\adaga$ and after the optical RWA reads
\begin{eqnarray}
\label{eq:hifirst}
H^I_1=\frac{\delta_m(t)}{2}\sigma_z&+\frac{\Omega_a}{2}\left[\sigma^+e^{i\eta_a(ae^{-i\nu t}+\adag e^{i\nu t})}+\textrm{H.c.} \right]+ \nonumber \\
&+\sum_{j}\frac{\Omega_j}{2}\left[\sigma^+e^{i\eta_j(ae^{-i\nu t}+\adag e^{i\nu t})}e^{i(\Delta_j t-\phi_j)}+\textrm{H.c.}\right],
\end{eqnarray}
where $\delta_m(t)$ follows an OU process and is responsible of the dephasing noise,  $\Delta_j=\omega_I-\omega_j$ is the detuning and $\eta_j$ the Lamb-Dicke of the $j$th laser. Note that the additional laser $a$ has zero detuning, $\Delta_a=0$ which ensures a carrier  interaction (i.e. a $\sigma_x$ proportional term) within the Lamb-Dicke regime, and when other terms, i.e. the ones with a linear dependence in the Lamb-Dicke parameter, can be averaged out  because of the condition $\Omega_a\eta \ll \nu$. Hence, only the first term of the following expansion is considered,
\begin{eqnarray}
\label{eq:exp_LD}
e^{i\eta(ae^{-i\nu t}+\adag e^{i\nu t})}=I&+i\eta\left(ae^{-i\nu t}+\adag e^{i\nu t} \right)-\nonumber\\
&-\frac{\eta^2}{2}\left(2\adaga+1+a^2e^{-2i\nu t}+(\adag)^2 e^{2i\nu t}\right)+\mathcal{O}(\eta^3).
\end{eqnarray}
In this way the additional continuous driving  $a$, provides a dressed spin-basis, $\left\{\upx,\downx\right\}$, in which the system  is protected against the magnetic-field fluctuation or dephasing noise, $\delta_m(t)/2 \sigma_z$, as long as $\Omega_a$  fulfills the criteria given in Sec.~\ref{sec:ccd}. Then, the magnetic-field fluctuation can be eliminated and the Hamiltonian~(\ref{eq:hifirst}) is
\begin{eqnarray}
\label{eq:H1Ieff}
H^{I}_1\approx \frac{\Omega_a}{2}\sigma_x +\sum_{j}\frac{\Omega_j}{2}\left[\sigma^+e^{i\eta_j(ae^{-i\nu t}+\adag e^{i\nu t})}e^{i(\Delta_j t-\phi_j)}+\textrm{H.c.}\right].
\end{eqnarray}
Furthermore, choosing properly the detunings and phases, $\Delta_j$ and $\phi_j$, a tunable Rabi model can be obtained from the previous effective Hamiltonian. This can be accomplished by setting two lasers $j=1,2$ with $\Delta_1=\nu-\xi$ and $\Delta_2=-\nu+\xi$ (detuned red and blue sideband), for which only the terms at first order in $\eta$ ($\eta_{1,2}=\eta$) of the expansion in Eq.~(\ref{eq:exp_LD}) survive, provided by $\xi\ll \nu$ and $\Omega_j\ll \nu$; that is, we are applying the vibrational RWA. Finally, the Rabi model is achieved when an interaction term is orthogonal to the free energy term of the two-level system, which in this case is $\sigma_x$. Therefore, it suffices to set the phases as $\phi_1=\phi_2=0$ and the Rabi frequencies $\Omega_{1,2}=\Omega$,
\begin{eqnarray}
H^{I}_1\approx \frac{\Omega_a}{2}\sigma_x -\frac{\Omega\eta}{2}\sigma_y\left(ae^{-i\xi t}+\adag e^{i\xi t} \right).
\end{eqnarray}
The previous Hamiltonian corresponds to a Rabi model in a rotating frame w.r.t. $\xi\adaga$, i.e.
\begin{eqnarray}
\label{eq:Hi_final1}
H_R= \frac{\Omega_a}{2}\sigma_x +\xi \adaga-\frac{\Omega\eta}{2}\sigma_y\left(a+\adag \right).
\end{eqnarray}
We remark that the previous effective Hamiltonian is only valid under both optical and  vibrational RWA, within the Lamb-Dicke regime and when $\Omega_a$ is such that the noise $\delta_m(t)$ has vanishing small component at that frequency.

 Under the same approximations, the Dirac equation can be obtained. The corresponding Hamiltonian  of the $(1+1)$ Dirac equation~\cite{Lamata07, Casanova10r} reads $H_D=c_D \hat{p} \sigma_x +m_D c^2 \sigma_z$, where $c_D$ is the speed of light, $m_D$ the mass of the $\frac{1}{2}$-spin particle, and $\hat{p}$ the momentum operator.  To realize such a Hamiltonian from Eq.~(\ref{eq:H1Ieff}), we  select $\Delta_1=\nu$, $\Delta_2=-\nu$ (red and blue sideband), $\phi_1=3\pi/2$, $\phi_2=\pi/2$ considering $\eta_{1,2}=\eta$ and $\Omega_{1,2}=\Omega$ (together with $\Delta_a=0$ and $\phi_a=0$). Then, Eq.~(\ref{eq:H1Ieff}) reads
\begin{eqnarray}
H_1^I\approx  \frac{\Omega_a}{2}\sigma_x +\eta\Omega\sigma_y\hat{p},
\end{eqnarray}
where $\hat{p}=i(\adag-a)/2$. This is equivalent to the Dirac equation with the following parameters $c_D=\eta\Omega$ and $m_D=\Omega_a/(2\eta^2\Omega^2)$.

\subsubsection{Second layer}
\label{subsubsec:2layer} 
Once the main source of noise, magnetic field fluctuations, is overcome by means of the first layer, the following step consists in facing laser-intensity fluctuations which can still spoil quantum coherence. The intensity of a $j$th laser is now modeled as $\Omega_j(t)=\Omega_j\left( 1+\delta_{\Omega_j}(t) \right)$, where $\Omega_j$ is the desired Rabi frequency and $\delta_{\Omega_j}(t)$ describes a small stochastic fluctuation. Such fluctuation will be present for all the lasers used in the setup. That is, the laser intensities are not completely stable, but fluctuate around its mean value $\Omega_j$. We characterize these fluctuations as an OU process with $\tau_{\Omega}=1$ ms following~\cite{Haffner08}, and an amplitude of $0.1\%$ ($p=0.001$) of the laser intensity $\Omega_j$. Thus, one can characterize this as $\sigma[\delta_\Omega]=p$, which leads to $c_{\Omega}=2p^2/\tau_{\Omega}$. Note that the laser-amplitude noise is chosen to be slow, compared to $\delta_m(t)$. This fact can be seen as a technological requirement as otherwise the noise might not be easily handled within the CCD scheme as we will discuss later on.

 In this way, once $\delta_m(t)/2 \ \sigma_z$ is overcome, the main fluctuation in Eq.~(\ref{eq:H1Ieff}) appears in the free energy term of the two-level system (i.e. as dephasing noise). Note that the rest of the Rabi frequencies, $\Omega_j$, are multiplied by a Lamb-Dicke parameter which reduces the influence of the errors introduced into the system by their fluctuating character. Therefore, we can proceed as for the first layer to deal with the  term $\Omega_a\delta_{\Omega_a}(t)/2 \ \sigma_x$. To eliminate its contribution an additional continuous driving, denoted by the subscript $b$, is introduced, but with a time-dependent Rabi frequency  $\Omega_b2\cos(\Omega_a t)$. The Hamiltonian describing this situation in a rotating frame w.r.t. $H_0=\omega_I/2 \sigma_z +\nu\adaga$ reads
\begin{eqnarray}
H^I_2\approx \frac{\delta_m(t)}{2}\sigma_z&+\frac{\Omega_a}{2}\sigma_x +\frac{\Omega_a\delta_{\Omega_a}(t)}{2}\sigma_x+\nonumber \\
&+\sum_{j}\frac{\Omega_j}{2}\left[\sigma^+e^{i\eta_j(ae^{-i\nu t}+\adag e^{i\nu t})}e^{i(\Delta_j t-\phi_j)}+\textrm{H.c.}\right]\nonumber \\
&+\frac{2\Omega_b\cos(\Omega_at)}{2}\left[\sigma^+e^{i\eta_b(ae^{-i\nu t}+\adag e^{i\nu t})}e^{-i\phi_b}\right],
\end{eqnarray}
where we have already fixed $\Delta_b=0$. By simplicity, we only write down explicitly the fluctuation $\delta_m(t)$ and $\delta_{\Omega_a}(t)$, although all the functions $\delta_{\Omega_j}(t)$ have been taken into account in our numerical simulations, see next Section. As we need an orthogonal carrier with respect to $\sigma_x$ for $\Omega_b$, we select  $\phi_b=\pi/2$ which leads to $\Omega_b\cos(\Omega_at)\sigma_y$. Now we move to a rotating frame w.r.t. $\Omega_a/2 \sigma_x$ obtaining
\begin{eqnarray}
H^{II}_2&=e^{i\frac{\Omega_a}{2}\sigma_xt}H^I_2e^{-i\frac{\Omega_a}{2}\sigma_xt}\nonumber \\&\approx\frac{\delta_m(t)}{2}\left[\cos(\Omega_at)\sigma_z+\sin(\Omega_at)\sigma_y\right]+\frac{\Omega_a\delta_{\Omega_a}(t)}{2}\sigma_x+\nonumber \\
&+\frac{\Omega_b}{2}\left[\cos^2(\Omega_at)\sigma_y-\cos(\Omega_at)\sin(\Omega_at)\sigma_z\right]+\nonumber\\&+\sum_j\frac{\Omega_j}{2}\left[e^{i\frac{\Omega_a}{2}\sigma_xt}\sigma^+e^{-i\frac{\Omega_a}{2}\sigma_xt} e^{i\eta_j(ae^{-i\nu t}+\adag e^{i\nu t})}e^{i(\Delta_j t-\phi_j)}+\textrm{H.c}\right].
\end{eqnarray}
The spin raising and lowering operators have contributions of $\sigma_x$ and $\sigma_y$, i.e. $\sigma^{\pm}=\frac{1}{2}(\sigma_x\pm i\sigma_y)$, which in a rotating frame with respect to $\Omega_a/2 \ \sigma_x$ makes $\sigma_y$ to rotate at frequencies $\pm\Omega_a$ while it does not affect $\sigma_x$. We then invoke the RWA to average out those rotating terms. Note that this is valid under the assumption $\Omega_b\ll\Omega_a$. The free energy term of the effective two-level system is given now by $\sigma_y$, and hence, the new dressed spin-basis is $\left\{\ket{\uparrow}_y,\ket{\downarrow}_y \right\}$. In this basis the fluctuating term $\Omega_a\delta_{\Omega_a}(t)/2 \ \sigma_x$ can be depreciated following the same arguments given in Sec.~\ref{sec:ccd}, as well as $\delta_m(t)$. Hence, the Hamiltonian can be  approximated by
\begin{eqnarray}
\label{eq:H2IIeff}
H_2^{II}\approx \frac{\Omega_b}{2}\sigma_y+\sum_j\frac{\Omega_j}{2}\left[\frac{\sigma_x}{2}e^{i\eta_j(ae^{-i\nu t}+\adag e^{i\nu t})}e^{i(\Delta_j t-\phi_j)}+\textrm{H.c}\right].
\end{eqnarray}
We can summarize the operating regime on the second layer as  $\Omega_b\ll\Omega_a\ll\omega_I$. Additionally, $\Omega_a$ has to be large enough to ensure decoupling with respect to $\delta_m(t)$, this condition is $\Omega_a\gtrsim 1/(2\pi\tau_m)$ or, in different words, $\Omega_a$ has to be larger than the crossover frequency, see Sec.~\ref{sec:OU}. At the same time, and following the same arguments, $\Omega_b$ needs to handle the fluctuation $\Omega_a\delta_{\Omega_a}(t)/2 \ \sigma_x$, and hence, $\Omega_b\gtrsim 1/(2\pi\tau_{\Omega})$ which implies the relation $\tau_{\Omega_a} \gg \tau_m$. Yet,  both the intensity of the noise  and the RWA ($\Omega_b\ll\Omega_a$) play a decisive role to successfully apply a second layer of protection in the CCD scheme.

We note that now we may use only one traveling wave to produce the Rabi-like interaction. Setting $\Delta_1=+\nu-\xi$, $\phi_1=3\pi/2$ , we arrive to 
\begin{eqnarray}
\label{eq:H2R}
H_2^{II}\approx \frac{\Omega_b}{2}\sigma_y-\frac{\Omega_1 \eta_1}{4}\sigma_x\left(ae^{-i\xi t}+\adag e^{i\xi t} \right),
\end{eqnarray}
after using the vibrational RWA. The previous equation is equivalent to the Rabi model in a rotating frame w.r.t. $\xi\adaga$,
\begin{eqnarray}
H_R=\frac{\Omega_b}{2}\sigma_y+\xi \adaga -\frac{\Omega \eta}{4}\sigma_x\left(a+\adag\right).
\end{eqnarray}
As in the case of the first layer, the Dirac equation can be realized in a straightforward manner. Choosing $\Omega_1=\Omega$, $\eta_1=\eta$, $\Delta_1=\nu$ and $\phi_1=\pi$ the Eq.~(\ref{eq:H2IIeff}) reduces to 
\begin{equation}
\label{eq:H2D}
H_2^{II}\approx \frac{\Omega_b}{2}\sigma_y+\frac{\eta \Omega}{2}\sigma_x\hat{p},
\end{equation}
which is equivalent to the Dirac Hamiltonian with $c_D=\eta\Omega/2$ and $m_D=2\Omega_b/(\eta^2\Omega^2)$. 
Note that the effective Hamiltonians given in Eqs.~(\ref{eq:H2R}) and~(\ref{eq:H2D}) are valid under a number of approximations, as for the first layer. Additionally, we now require $\Omega_b\ll\Omega_a$ due to a RWA, but at the same time $\Omega_b$ must be still large enough to decouple with respect to the noisy term $\Omega_a\delta_{\Omega_a}(t)\sigma_x$.


\section{Numerical results}
\label{sec:num}
Here we present numerical simulations of the previous derived effective Hamiltonians. We compare the usefulness of CCD scheme in contrast to the bare realization, denoted  here as {\em zeroth layer} (see for example~\cite{Pedernales:15} and~\ref{ap:1} for a derivation), i.e., when no protection against noise is provided. We explore two physical regimes in the realized quantum Rabi model, namely, the paradigmatic resonant case to observe Rabi oscillations, and the limiting case where a quantum phase transition takes place~\cite{Hwang:15,Puebla:16}. Then, we present the case of the evolution of a Dirac particle. We emphasize that all the numerical simulations involving trapped-ion Hamiltonians have been carried out after the optical RWA without performing further approximations. 

The bare realization or {\em zeroth layer} is accomplished by two lasers
\begin{eqnarray}
\label{eq:H0sim}
H_0^I=\frac{\delta_m(t)}{2}\sigma_z&+\frac{\Omega_1(1+\delta_{\Omega_1}(t))}{2}\left[\sigma^+e^{i\eta_1\left(ae^{-i\nu t}+\adag e^{i\nu t} \right)}e^{i(\Delta_1t-\phi_1)}+ \textrm{H.c.}\right]+\nonumber \\&+\frac{\Omega_2(1+\delta_{\Omega_2}(t))}{2}\left[\sigma^+e^{i\eta_2\left(ae^{-i\nu t}+\adag e^{i\nu t} \right)}e^{i(\Delta_2t-\phi_2)}+ \textrm{H.c.}\right],\nonumber\\
\end{eqnarray}
while the first layer involves and additional laser for protection purposes,
\begin{eqnarray}
\label{eq:H1sim}
H_1^I=\frac{\delta_m(t)}{2}\sigma_z&+\frac{\Omega_1(1+\delta_{\Omega_1}(t))}{2}\left[\sigma^+e^{i\eta_1\left(ae^{-i\nu t}+\adag e^{i\nu t} \right)}e^{i(\Delta_1t-\phi_1)}+ \textrm{H.c.}\right]\nonumber \\
&+\frac{\Omega_2(1+\delta_{\Omega_2}(t))}{2}\left[\sigma^+e^{i\eta_2\left(ae^{-i\nu t}+\adag e^{i\nu t} \right)}e^{i(\Delta_2t-\phi_2)}+ \textrm{H.c.}\right]\nonumber\\
&+\frac{\Omega_a(1+\delta_{\Omega_a}(t))}{2}\left[\sigma^+e^{i\eta_a\left(ae^{-i\nu t}+\adag e^{i\nu t} \right)}e^{i(\Delta_at-\phi_a)}+ \textrm{H.c.}\right].\nonumber\\
\end{eqnarray}
Finally, the second layer adds a time-dependent Rabi frequency, 
\begin{eqnarray}
\label{eq:H2sim}
H_2^{I}=\frac{\delta_m(t)}{2}\sigma_z&+\frac{\Omega_1(1+\delta_{\Omega_1}(t))}{2}\left[\sigma^+e^{i\eta_1\left(ae^{-i\nu t}+\adag e^{i\nu t} \right)}e^{i(\Delta_1t-\phi_1)}+ \textrm{H.c.}\right]\nonumber \\
&+\frac{\Omega_a(1+\delta_{\Omega_a}(t))}{2}\left[\sigma^+e^{i\eta_a\left(ae^{-i\nu t}+\adag e^{i\nu t} \right)}e^{i(\Delta_at-\phi_a)}+ \textrm{H.c.}\right]\nonumber\\
&+\frac{2\Omega_b\cos(\Omega_at)(1+\delta_{\Omega_b}(t))}{2}\left[\sigma^+e^{i\eta_b\left(ae^{-i\nu t}+\adag e^{i\nu t} \right)}e^{i(\Delta_bt-\phi_b)}+ \textrm{H.c.}\right].\nonumber\\
\end{eqnarray}
The effective magnetic-field fluctuation is described by $\delta_m(t)$, as shown in Sec.~\ref{sec:OU} and~\ref{sec:ccd}, with parameters $\tau_m=50 \ \mu$s and $T_2=3$ ms. Note that distinct experimental setups may suffer different magnetic-field fluctuation, and thus $\tau_m$ may differ. In this respect, depending on the correlation time $\tau_m$, our scheme can be adapted to suppress magnetic-field fluctuations by setting properly the Rabi frequencies $\Omega_j$, as discussed in Sec.~\ref{sec:ccd}. However, for a too short noise correlation time, i.e. in the limit of Markovian noise $\tau_m/T_2\rightarrow 0$, the tunability of the simulated Rabi models using CCD scheme is reduced as the Rabi frequency must fulfill $\Omega_a>1/(2\pi\tau_m)$ to ensure decoupling. We recall that the characteristic frequency from which the spectral density starts to decay as $1/f^2$ corresponds to $f_{cr}=1/(2\pi\tau_m)$, and therefore $\Omega_a>f_{cr}$, as explained in Sec.~\ref{sec:ccd}. In addition, the fluctuation of the  $j$th laser's amplitude, denoted as $\delta_{\Omega_j}(t)$, is parametrized with $\tau_\Omega=1$ ms and $c_\Omega=2p^2/\tau_\Omega$ as it describes a relative amplitude fluctuation, with $p=0.1\%$. We have considered an equal noise for the lasers with intensities $\Omega_1$ and $\Omega_2$, i.e. $\delta_{\Omega_1}(t)=\delta_{\Omega_2}(t)$, while the fluctuations of the rest are completely independent. However, we also performed simulations with uncorrelated noise between $\Omega_1$ and $\Omega_2$ and no significant differences have been observed. In all the simulations, the trap frequency has been chosen as $\nu=2\pi\times1.36 \ \MHz$, the Lamb-Dicke parameter as $\eta_{1,2}=0.06$ and $\eta_{a,b}=0.01$~\cite{Gerritsma09,Gerritsma11}.

\subsection{Quantum Rabi model realization}
\label{subsec:QRM}
Here we present the numerical simulations of the trapped-ion Hamiltonian realizing the quantum Rabi model to observe the paradigmatic Rabi oscillations. The simulated quantum Rabi model in the $i$th layer can be written as
\begin{equation}
\label{eq:simR}
  H_{R,i}=\frac{\tilde{\Omega}_i}{2}\sigma^i_{\tiny{\textrm{TLS}}}+\tilde{\omega}_i\adaga-\tilde{\lambda}_i\sigma_{\perp}^i\left(a+\adag\right),
\end{equation}
where $\sigma_{\tiny{\textrm{TLS}}}^i$ and $\sigma_{\perp}^i$ stand for the Pauli matrices of the free energy term of the two-level system and the orthogonal direction of the interaction, respectively. The parameters used to simulate this model using Eqs.~(\ref{eq:H0sim}),~(\ref{eq:H1sim}) and~(\ref{eq:H2sim}) are gathered in Table~\ref{tab:1}, as well as their relation with the effective frequencies given in Eq.~(\ref{eq:simR}), $\tilde{\Omega}_i$, $\tilde{\omega}_i$ and $\tilde{\lambda}_i$. Note that $\Omega_{1,2}=\Omega$ and $\eta_{1,2}=\eta$ for zeroth and first layer.
\begin{table}
\caption{\label{tab:1} Trapped-ion parameters to simulate the quantum Rabi model using CCD scheme.}
\begin{indented}
\item[]\begin{tabular}{@{}llll}
\br
${}$&Zeroth layer&First layer&Second layer\\
\mr
$\Delta_1$ & $\nu+\delta_1$& $\nu-\omega_1$&$\nu-\omega_2$\\
$\Delta_2$ & $-\nu+\delta_2$& $-\nu+\omega_1$& ---\\
$\Delta_a$ & --- & $0$ & 0\\
$\Delta_b$ & --- & --- & 0\\
$\phi_{1,2}$& $3\pi/2$ & $3\pi/2$ & $3\pi/2$ \\
$\phi_a$ & --- & $0$ & 0\\
$\phi_b$ & --- & --- & $\pi/2$\\
\mr
$\sigma_{\tiny{\textrm{TLS}}}^i$ & $\sigma_z$ & $\sigma_x$ & $\sigma_y$  \\
$\sigma_{\perp}^i$ & $\sigma_x$ & $\sigma_y$ & $\sigma_x$  \\
$\tilde{\Omega}_i$ &$\frac{1}{2}(\delta_2+\delta_1)$ & $\Omega_a$ & $\Omega_b$ \\
$\tilde{\omega}_i$ & $\frac{1}{2}(\delta_2-\delta_1)$ & $\omega_1$ & $\omega_2$  \\
$\tilde{\lambda}_i$ & $\frac{\eta\Omega}{2}$ & $\frac{\eta\Omega}{2}$ & $\frac{\eta_1\Omega_1}{4}$\\
\br
\end{tabular}
\end{indented}
\end{table}
In order to achieve the same effective model, regardless of the layer, we will introduce dimensionless constants to define a target Hamiltonian. These are $R\equiv\tilde{\Omega}_i/\tilde{\omega}_i$ and $g\equiv2\tilde{\lambda}_i/(\tilde{\omega}_i\sqrt{R})$. Hence, fixing $R$ and $g$, $H_{R,i}/\tilde{\omega}_i$ represents the same effective quantum Rabi model.
 \begin{figure}
\centering
\includegraphics[width=0.45\linewidth,angle=-90]{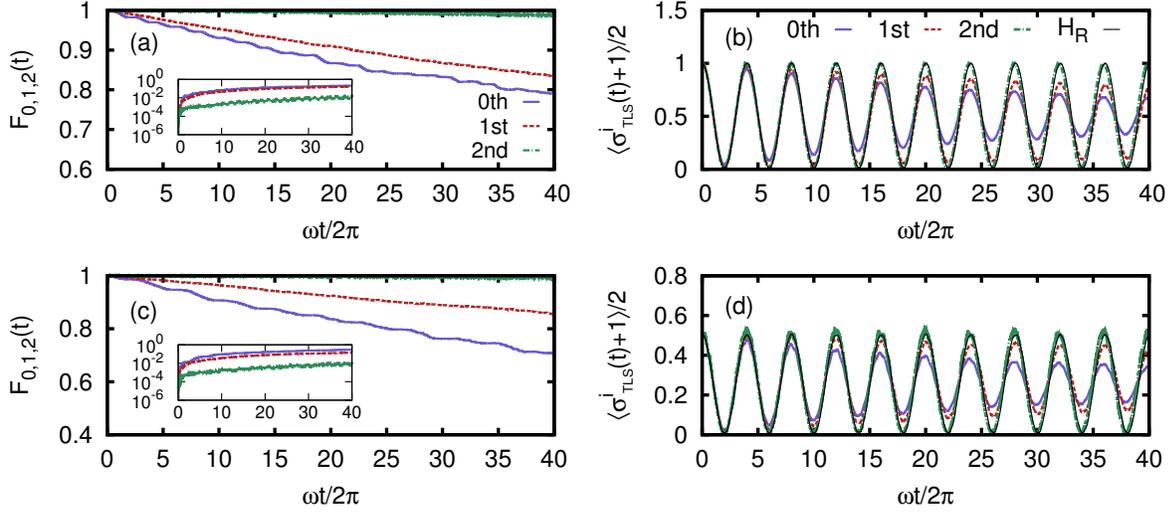}
\caption{{\small Trapped-ion realization of the quantum Rabi model with different levels of protection. Time evolution of the population of the two-level system in $H_{0,1,2}$ (b) and (d), and the fidelity $F_{0,1,2}(t)$ (a) and (c), where the infidelity $1-F_{0,1,2}(t)$ is plotted in the inset. In (a) and (b) the initial state is $\ket{\psi(0)}=\ket{0}\ket{\uparrow}_{\tiny{\textrm{TLS}}}$, while in (c) and (d) $\ket{\psi(0)}=\ket{0}\ket{\uparrow}_{\perp}$. The black line gives the ideal, noiseless, case of $H_R$ in (b) and (d). The results were obtained averaging over $200$ stochastic trajectories; solid (light blue) line, dashed (red) line and dot-dashed (green) line correspond to zeroth, first and second layer, respectively. As $\tilde{\omega}_{0,1,2}=2\pi\times5 \ \kHz$, the total evolution time corresponds to $8$ ms. See main text for simulation parameters. }}
\label{fig:TIRabiCCD}
\end{figure}
We set $\tilde{\omega}_{0,1,2}=\tilde{\Omega}_{0,1,2}=2\pi\times 5 \ \kHz$ to simulate a resonant case $R=1$, and a dimensionless coupling constant $g=1/4$. This implies that: (i) for $H_0^I$, i.e. for the bare realization, $\delta_2=2\pi\times10 \ \kHz$, $\delta_1=0$ and $\Omega_{1,2}=2\pi\times20.83 \ \kHz$; (ii) for $H_1^I$ (first layer) $\omega_1=2\pi\times 5 \ \kHz$, $\Omega_a=2\pi\times 5 \ \kHz$ and $\Omega_{1,2}=2\pi\times20.83 \ \kHz$; (iii) for $H_2^I$ (second layer) $\omega_2=2\pi\times 5 \ \kHz$, $\Omega_b=2\pi\times 5 \ \kHz$ and $\Omega_a=40\Omega_b=2\pi\times200 \ \kHz$, $\Omega_1=2\pi\times41.67 \ \kHz$. 

We illustrate how CCD improves the realization of the Rabi model by means of the fidelity among the wavefunction of the ideal Rabi model, $\ket{\psi_{R,i}(t)}$, and its noisy trapped-ion realization $\ket{\psi_i(t)}$ for the $i$th layer of protection, which reads
\begin{eqnarray}
F_i(t)=\left| \left< \psi_{R,i}(t)\right| \left.\psi_i(t) \right> \right|.
\end{eqnarray} 
We will also compare  the oscillations of the population on the excited state of the qubit which is given by $\braket{\sigma_{\tiny{\textrm{TLS}}}^i+1}/2$ in both cases, ideal and the trapped ion realization with different noisy contributions.

\begin{figure}
\centering
\includegraphics[width=0.25\linewidth,angle=-90]{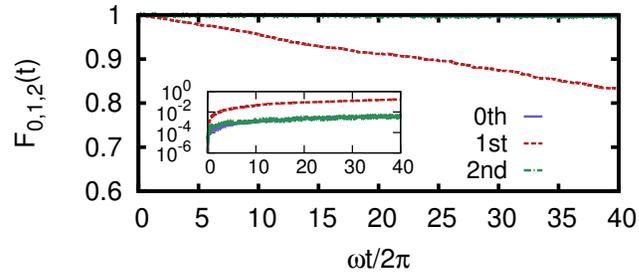}
\caption{{\small Trapped-ion realization of the quantum Rabi model.  The results were obtained averaging over $200$ stochastic trajectories. The solid (light blue) line, dashed (red) line and dot-dashed (green) line correspond to zeroth, first and second layer, respectively. Note that in this case the first layer deteriorates the fidelity, which is a consequence of the particular state and parameters considered, since $\ket{\psi(0)}=\ket{0}\ket{\downarrow}_{\tiny{\textrm{TLS}}}$ is a dark state (see main text for further details). In the inset the infidelity $1-F_{0,1,2}(t)$ is plotted. Note that the results for the zeroth and second layer completely overlap. The final time corresponds to $8$ ms. See main text for simulation parameters.}}
\label{fig:TIRabiCCD_SdownZ_F}
\end{figure}
In Figs.~\ref{fig:TIRabiCCD} the improvement achieved by applying the CCD scheme is clearly demonstrated for two different initial states, $\ket{\psi(0)}=\ket{0}\ket{\uparrow}_{\tiny{\textrm{TLS}}}$ and $\ket{\psi(0)}=\ket{0}\ket{\uparrow}_{\perp}$, where $\sigma_{\tiny{\textrm{TLS}}}\ket{\uparrow}_{\tiny{\textrm{TLS}}}=+\ket{\uparrow}_{\tiny{\textrm{TLS}}}$ and $\sigma_{\perp}\ket{\uparrow}_{\perp}=+\ket{\uparrow}_{\perp}$. To the contrary, there are specific situations in which CCD scheme could deteriorate the desired realization. In particular, if the considered initial state is parallel to both magnetic noise, $\delta_m\sigma_z$ and Hamiltonian (i.e. when we deal with the dark state), to apply CCD scheme is counterproductive since it changes a source of noise, that originally just gives rise to a global phase, to an orthogonal noise producing transitions and distorting the dynamics. This is the case for $\ket{\psi(0)}=\ket{0}\ket{\downarrow}_{\tiny{\textrm{TLS}}}$ in the Rabi model when $g\ll 1$ i.e. when the Jaynes-Cummings model arises. As we see in Fig.~\ref{fig:TIRabiCCD_SdownZ_F}, for $R=1$ and $g=1/4$ the fidelity of the first layer is noticeably worse that an unprotected realization, while the second layer is just as good as the original. This reveals that CCD scheme does not necessarily lead to an improved realization; it depends on several factors which have to be taken into account beforehand.

\subsection{Critical dynamics of the superradiant quantum phase transition in the Rabi model}
In order to illustrate the versatility of the CCD scheme, we analyze the realization of a time-dependent Rabi Hamiltonian in the ultra-strong coupling regime. In this respect, it has been recently shown that the Rabi model (Eq.~(\ref{eq:simR})) undergoes a quantum phase transition in the $R=\Omega/\omega_0\rightarrow\infty$ limit at the critical point $g_c=2\lambda_c/\sqrt{\Omega\omega_0}=1$ despite of consisting only of a single two-level system and a single-mode bosonic field~\cite{Hwang:15}. For finite $R$,  critical behavior is revealed in the form of \textit{finite-frequency} scaling functions, in an approach that is equivalent to finite-size scaling in traditional phase transitions~\cite{Fisher:72,Botet:82}. As shown in~\cite{Puebla:16}, the presence of the quantum phase transition can be observed with a single trapped-ion that interacts with one of its vibrational modes. This can be achieved resorting to non-equilibrium universal scaling functions~\cite{Acevedo:14,Puebla:16} in terms of the expectation value $\left<\sigma^i_{\tiny{\textrm{TLS}}}\right>$ of Eq.~(\ref{eq:simR}), which can be measured with high-fidelity in a trapped-ion system~\cite{Myerson08, Burrell10}. To obtain such non-equilibrium universal scaling functions one can proceed as follows. Prepare an initial state $\ket{\psi(0)}=\ket{0}\ket{\downarrow}_{\tiny{\textrm{TLS}}}$ at $g=0$ for a fixed $R$, such that $\sigma_{\tiny{\textrm{TLS}}}^i\ket{\downarrow}_{\tiny{\textrm{TLS}}}=-\ket{\downarrow}_{\tiny{\textrm{TLS}}}$, and then quench continuously in a time $\tau_Q$ the coupling constant $g$ until $g=g_c=1$ is reached. Then, at $g(\tau_Q)=1$ for a frequency ratio $R$ we calculate the quantity $\braket{\sigma_{\tiny{\textrm{TLS}}}^i}_R(\tau_Q,R)=\left|\left<\psi(\tau_Q)\left|\sigma_{\tiny{\textrm{TLS}}}^i \right|\psi(\tau_Q)\right>-\braket{\sigma{\tiny{\textrm{TLS}}}^i}_{GS}(R) \right|$, where $\braket{\sigma{\tiny{\textrm{TLS}}}^i}_{GS}(R)$ is the ground-state expectation value of $\sigma_{\tiny{\textrm{TLS}}}^i$ at $g=1$ and $R$. The non-equilibrium universal function is found as $S(T)=R^{\mu}\braket{\sigma_{\tiny{\textrm{TLS}}}^i}_R$ where $T\equiv R^{-\gamma/(\mu(1+\zeta))}\tau_Q$. The critical exponents are $\mu=2/3$, $\gamma=1$ and $\zeta=1/2$~\cite{Hwang:15,Puebla:16}. Note however that the driving time $\tau_Q$ cannot be arbitrarily short since $S(T)$ is obtained assuming adiabatic dynamics away from the critical point. On the other hand, in an ion-trap realization, the duration of the dynamics to reconstruct $S(T)$ is severely restricted due to the presence of various sources of noise~\cite{Puebla:16}.

Here, by applying the CCD scheme, we offer a way to overcome these noises, which facilitates the observation of universal scaling functions, and illustrate that the CCD scheme is valid in an extreme parameter regime and even when quench dynamics is considered. Note however that, due to the large desired value of $R$, the second layer is expected to fail as $R\propto \Omega_b$ but $\Omega_b\ll\Omega_a$ is required to fulfill the RWA. Hence, for this specific case the approximations leading to the quantum Rabi model will break down.

\begin{figure}
\centering
\includegraphics[width=0.3\linewidth,angle=-90]{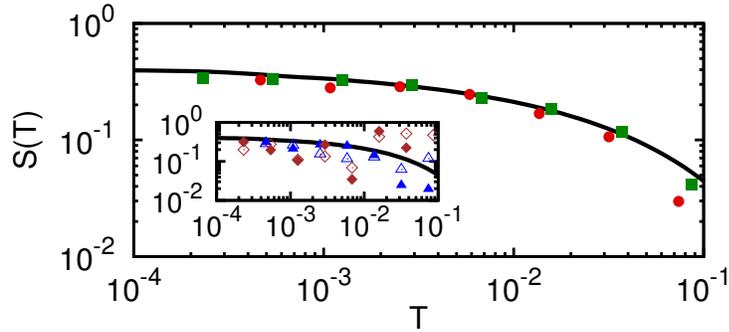}
\caption{{\small Realization of the universal non-equilibrium function $S(T)$ of the Rabi model using a trapped-ion setting. The points correspond to the simulated trapped-ion Hamiltonian using one layer of protection, for $R=50$ (circles) and $R=100$ (squares). Each point has been obtained averaging over $100$ stochastic trajectories. In the inset we represent the obtained $S(T)$ without protection (open symbols) and in the second layer (full symbols), which does not show the expected collapse as a consequence of noises and breakdown of the approximations. The driving time $\tau_Q$ ranges from $0.02$ ms to $8.6$ ms for zeroth and first layer, and from $0.05$ ms to $9.23$ ms for the second layer. See main text for further details.}}
\label{fig:TIQPTuniv}
\end{figure}

The Fig.~\ref{fig:TIQPTuniv} shows the universal non-equilibrium function $S(T)$ as a function of the rescaled driving time $T$. The solid black line corresponds to the ideal quantum Rabi model, while the points to the trapped-ion realization using a first layer protection with $R=50$ (circles) and $R=100$ (squares) for $0.02\leq\tau_Q\leq 8.6$ in units of $2\pi/\tilde{\omega}_i$. In the inset the results using zeroth and second layer are plotted. Observe the remarkable improvement compared to the zeroth layer, and the failure of the second layer as $\Omega_b$ becomes comparable to $\Omega_a$. The simulation parameters are $\tilde{\omega}_{0,1}=2\pi\times 1 \ \kHz$, $\tilde{\omega}_2=2\pi\times400\Hz$, while $\tilde{\Omega}_i=R\tilde{\omega}_i$. For the second layer $\Omega_{a}$ is set to $2\pi\times200 \ \kHz$, and hence $\Omega_a/\Omega_b=10$ and $5$ for $R=50$ and $100$, respectively, which already provides evidence of the expected failure of the RWA. Additionally, the quench is attained by tuning linearly in time the laser intensities from $0$ to $\Omega_f$. For the zeroth and first layer, $\Omega_f$ results in $2\pi\times 117.8 \ \kHz$ and $2\pi\times 166.7 \ \kHz$ for $R=50$ and $R=100$, respectively. For the second layer $\Omega_f$ amounts to $2\pi\times 94.3 \ \kHz$ and $2\pi\times 133.3 \ \kHz$ for $R=50$ and $R=100$, respectively.
 
\begin{figure}
\centering
\includegraphics[width=0.225\linewidth,angle=-90]{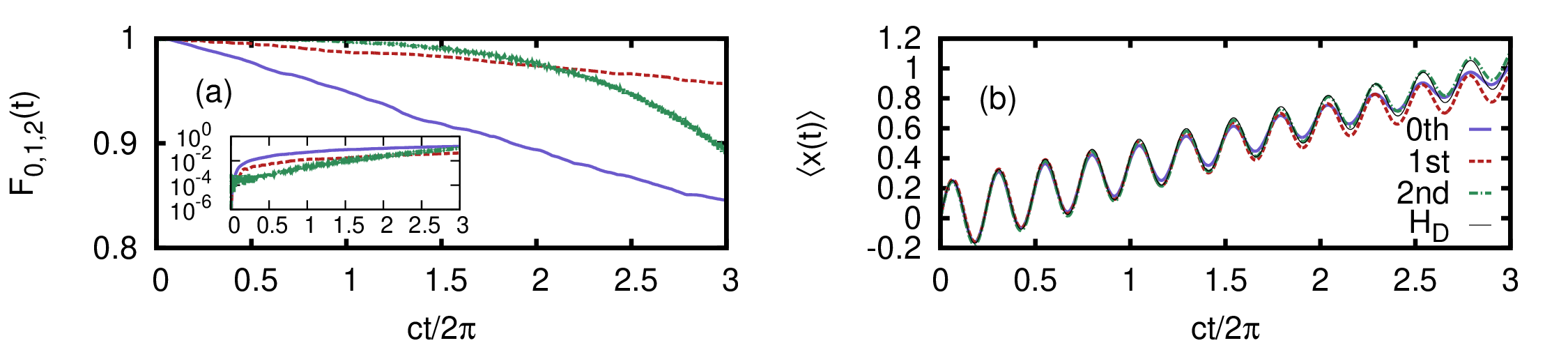}
\caption{{\small Trapped-ion realization of the Dirac equation. Time evolution of the fidelity (a) and position operator $\braket{x(t)}$ (b), with an initial state $\ket{\psi(0)}=\ket{0}\ket{\uparrow}_{\perp}$ and $r=2$. In (a), the inset corresponds to the infidelity $1-F_{0,1,2}(t)$. The results were obtained averaging over $200$ stochastic trajectories; solid (light blue) line, dashed (red) line and dot-dashed (green) line correspond to zeroth, first and second layer, respectively. In (b), solid black line corresponds to the ideal, noiseless, case of $H_D$. The total evolution time corresponds to $2.4$ ms. See main text for simulation parameters.}}
\label{fig:TIDiracCCD}
\end{figure}

\subsection{Dirac equation realization in a trapped-ion setting}
The parameters to realize the Dirac equation, $H_{D,i}/c_D=r\sigma_{\tiny{\textrm{TLS}}}^i+\hat{p}\sigma_{\perp}^i$ with $r\equiv m_Dc_D$, using Eqs.~(\ref{eq:H0sim}),~(\ref{eq:H1sim}) and~(\ref{eq:H2sim}) are gathered in the Table~\ref{tab:2}. 
\begin{table}
\caption{\label{tab:2} Trapped-ion parameters to simulate the Dirac equation using CCD scheme.}
\begin{indented}
\item[]\begin{tabular}{@{}llll}
\br
${}$&Zeroth layer&First layer&Second layer\\
\mr
$\Delta_1$ & $\nu+\delta$& $\nu$&$\nu$\\
$\Delta_2$ & $-\nu+\delta$& $-\nu$ & ---\\
$\Delta_a$ & --- & $0$ & $0$\\
$\Delta_b$ & --- & --- & $0$\\
$\phi_{1}$& $\pi$ & $3\pi/2$ & $\pi$ \\
$\phi_2$ & $0$ & $\pi/2$ & --- \\
$\phi_a$ & --- & $0$ & $0$\\
$\phi_b$ & --- & --- & $\pi/2$\\
\mr
$\sigma_{\tiny{\textrm{TLS}}}^i$ & $\sigma_z$ & $\sigma_x$ & $\sigma_y$  \\
$\sigma_{\perp}^i$ & $\sigma_x$ & $\sigma_y$ & $\sigma_x$  \\
$m_Dc_D^2$ &$\frac{\delta}{2}$ & $\frac{\Omega_a}{2}$ & $\frac{\Omega_b}{2}$ \\
$c_D$ & $\eta\Omega$ & $\eta\Omega$ & $\frac{\eta_1\Omega_1}{2}$ \\
\br
\end{tabular}
\end{indented}
\end{table}

In order to observe the paradigmatic Zitterbewegung~\cite{Gerritsma09} we calculate the expectation value of  the position operator $\hat{x}=(a+\adag)$ as a function of time for an initial state $\ket{\psi(0)}$, eigenstate of $\sigma_{\perp}^i$ (in particular we consider $\ket{\uparrow}_{\perp}$). We then set a value $m_D$ and $c_D$, or equivalently, $r$. Note that the presented scheme for first and second layer does not allow for a realization of the strict massless limit, $r=0$, since $r$ is proportional to $\Omega_{a}$ or $\Omega_b$ and $\Omega_{a,b}=0$ does not provide a protected Hamiltonian against fluctuations, while in the zeroth layer, $r$ is just proportional to the detuning $\delta$. Nevertheless, for $r>0$, CCD scheme still improves the simulated Dirac equation, as we illustrate in the following.

We set $r=2$, (i) $\delta=2\pi\times5 \ \kHz$, (ii) $\Omega_a=2\pi\times5 \ \kHz$, (iii) $\Omega_b=2\pi\times5 \ \kHz$ and $\Omega_a=2\pi\times200 \ \kHz$.  This implies (i) for Eq.~(\ref{eq:H0sim}) $\Omega_{1,2}=2\pi\times20.8 \ \kHz$ and $\Delta_{1,2}=\pm\nu+\delta$, (ii) for Eq.~(\ref{eq:H1sim}) $\Omega_{1,2}=2\pi\times 20.8 \ \kHz$ and (iii) for Eq.~(\ref{eq:H2sim}) $\Omega_1=2\pi\times41.7 \ \kHz$. In Fig.~\ref{fig:TIDiracCCD} we plot the fidelity  $F_{0,1,2}(t)$ (a) and position expectation value $\braket{x(t)}$ (b) as a function of time. The fidelity corresponds to $F_i(t)=\left| \left<\psi_{D,i}(t)\right| \left.\psi_{i}(t) \right>\right|$, where $\ket{\psi_i(t)}$ and $\ket{\psi_{D,i}(t)}$ are the wave-function of the trapped-ion and ideal Dirac equation of the $i$th layer, respectively. Note that the final time corresponds to $t=3(2\pi/c_D)=2.4$ ms. The improvement is clearly shown in Fig.~\ref{fig:TIDiracCCD}. The second layer works worse at longer times than the first one, which is mainly due to laser-amplitude fluctuations  and breakdown of RWA (note that $\Omega_a=40\Omega_b$). Nevertheless, for shorter times, the simulation of Dirac equation in the second layer is considerably enhanced. Finally we want to comment that the access to motional variables is achieved by, for example, adding a second ion to the trap and  computing the time derivative of the qubit expectation value~\cite{Gerritsma09, Gerritsma11}, see \ref{ap:2} for more details.
In principle, this protocol requires to prepare the ancillary ion in a certain quantum state that we will select as parallel to the magnetic noise $\delta_m(t)$. Hence, during the realization of the dynamics, this ion is not affected by external fluctuations, while, for the reconstruction of the time derivatives, a fast evolution is required. In this manner  the noise will have an small incidence in the reconstruction of $\langle x(t)\rangle$.

\section{Summary}
\label{sec:conc}
In the present article we demonstrate that concatenated continuous dynamical decoupling (CCD) can be applied to  a trapped-ion setup for a robust realization of the quantum Rabi model. We show that the use of the CCD scheme can significantly  improve the  coherence times and fidelities of quantum simulations in ion-trap experiments.  We exemplify this by means of numerical simulations exploiting the rich physics of the quantum Rabi model in three completely different parameter regimes.

\ack
This work is supported by an Alexander von Humboldt Professorship, the EU STREP project EQUAM, the ERC Synergy grant BioQ and the CRC TRR21. 
The authors acknowledge support by the state of Baden-W\"urttemberg through bwHPC
and the Germany Research Foundation (DFG) through grant no INST 40/467-1 FUGG. J. C. acknowledges support to the  Alexander von Humboldt foundation.

\appendix
\section{Zeroth layer realization of the quantum Rabi model}
\label{ap:1}
Here we recall briefly the procedure to realize the Rabi model and the Dirac equation without resorting to CCD scheme, as shown in~\cite{Pedernales:15}.

A tunable quantum Rabi model can be realized as follows. The trapped-ion Hamiltonian, in the rotating frame with respect to $\omega_I/2\sigma_z+\nu\adaga$ and after the optical RWA, reads
\begin{eqnarray}
\label{eq:H0I}
H_0^I=&\frac{\delta_m(t)}{2}\sigma_z+\frac{\Omega_1}{2}\left[\sigma^+e^{i\eta_1\left(ae^{-i\nu t}+\adag e^{i\nu t} \right)}e^{i(\Delta_1t-\phi_1)}+ \textrm{H.c.}\right]\nonumber \\
&+\frac{\Omega_2}{2}\left[\sigma^+e^{i\eta_2\left(ae^{-i\nu t}+\adag e^{i\nu t} \right)}e^{i(\Delta_2t-\phi_2)}+ \textrm{H.c.}\right].
\end{eqnarray}
Now, choosing frequency detunings such that $\Delta_1=\nu+\delta_1$ $\Delta_2=-\nu+\delta_2$, together with $\Omega_{1,2}=\Omega$, $\eta_{1,2}=\eta$ and $\phi_{1,2}=3\pi/2$ we obtain
\begin{eqnarray}
H_0^I&=\frac{\delta_m(t)}{2}\sigma_z-\frac{\eta\Omega}{2}\left[\sigma^+\left(ae^{i\delta_1t}+\adag e^{i\delta_2t}\right) +\textrm{H.c.}\right]\\
&=\frac{\delta_m(t)}{2}\sigma_z-\frac{\eta\Omega}{2}\left[(\sigma^+e^{i\tilde{\Omega}_0t}+\sigma^-e^{-i\tilde{\Omega}_0t})(ae^{-i\tilde{\omega}_0t}+\adag e^{i\tilde{\omega}_0t}) \right],
\end{eqnarray}
which corresponds to a Rabi model in a rotating frame  with respect to $\tilde{\Omega}_0/2\sigma_z+\tilde{\omega}_0\adaga$, being $\tilde{\Omega}_0=(\delta_1+\delta_2)/2$ and $\tilde{\omega}_0=(\delta_2-\delta_1)/2$.

In a straightforward manner, the Dirac equation is realized when choosing $\delta_{1,2}=\delta$, $\phi_1=\pi$, $\phi_2=0$, $\eta_{1,2}=\eta$ and $\Omega_{1,2}=\Omega$. Then, the Eq.~(\ref{eq:H0I}) adopts the following form
\begin{eqnarray}
H_0^I\approx \frac{\delta_m(t)}{2}\sigma_z+\eta\Omega\left[\sigma^+e^{i\delta t}+\sigma^-e^{-i\delta t}\right]\hat{p},
\end{eqnarray}
where $\hat{p}=i(\adag-a)/2$. The previous Hamiltonian is then equivalent to the Dirac Hamiltonian $H_D=\frac{\delta}{2}\sigma_z+\eta\Omega\sigma_x\hat{p}$ in a rotating frame with respect to $\delta/2\sigma_z$ (omitting fluctuations). Thus, $c_D=\eta\Omega$ and $m_Dc^2=\delta/2$.

\section{Measurement of vibrational operators}
\label{ap:2}

After the system evolution within the CCD scheme we have that the final state is $|\psi(t')\rangle$. Then, we can use another ion which is initialized into the state $\up$, and therefore does not suffer from the action of a noisy term $\delta_m(t)/2 \sigma_z^A$, where $\sigma_i^A$ are the Pauli operators of the ancillary ion. Hence, it does not require CCD protection. After the final time $t'$, a short evolution of time $t$ of the form $U = e^{-i\Omega t \sigma_x^A \hat{x}}$ is applied to the state  $\ket{\psi(t')}\up$. Then, it is easy to demonstrate that
\begin{equation}
\partial_t \langle \sigma_ y^A \rangle\bigg|_{t=0} = 2\Omega \langle \psi(t') | \hat{x} |  \psi(t') \rangle.
\end{equation}
\section*{References}
\bibliographystyle{iopart-num.bst}

\end{document}